\documentclass[aps,showpacs,preprintnumbers,nofootinbib]{revtex4}%
\usepackage{amssymb}
\usepackage{amsmath}
\usepackage{bm}
\usepackage{amsfonts}
\usepackage{graphicx}%
\providecommand{\U}[1]{\protect\rule{.1in}{.1in}}
\begin{document}
\title{Russell on Weyl's unified field theory }
\author{C. Romero }
\affiliation{Departamento de F\'{\i}sica, Universidade Federal da Para\'{\i}ba, Caixa
Postal 5008, 58059-970 Jo\~{a}o Pessoa, PB, Brazil }
\affiliation{E-mail: cromero@fisica.ufpb.br}

\begin{abstract}
In 1918, H. Weyl proposed a unified theory of gravity and electromagnetism
based on a generalization of Riemannian geometry. With hindsight we now could
say that the theory carried with it some of the most original ideas that
inspired the physics of the twentieth century. In a book published in 1927,
Bertrand Russell devoted an entire chapter to explain and give a critical
appraisal of Weyl%
\'{}%
s theory. We briefly revisit the text written by Russell, who gave one of the
first philosophical approaches to Weyl's ideas.

\end{abstract}

\pacs{04.20.Jb, 11.10.kk, 98.80.Cq}
\maketitle

\vskip .5cm

Keywords: Weyl's unified theory

\section{Introduction}

Weyl's attempt at unifying gravity and electromagnetism came to light in 1918
\cite{Weyl}. With hindsight we now could say that the theory carried with it
some of the most original ideas that inspired the physics of the twentieth
century. Indeed, it gave us a clue of how is possible, by extending already
known geometrical frameworks, to achieve unification of the gravitational and
non-gravitational physical fields, the latter being viewed as fundamental
ingredients of the spacetime geometry. In addition to this innovative result,
his search for hidden symmetries of space-time, somehow implied by the new
geometry contained the germs of the modern gauge theories of elementary
particles\cite{Straumann} .

\qquad Although Weyl's theory was not considered by Einstein to constitute a
viable physical theory, the powerful and elegant ideas put forward by the
publication of Weyl's paper survived and now constitutes a constant source of
inspiration for new proposals, particularly in the domain of the so-called
\textquotedblleft modified gravity theories\textquotedblright\ \cite{Sanomiya}.

\qquad Despite Einstein's objections, Weyl's unified theory attracted the
attention of some eminent contemporary physicists of Weyl, among whom we can
quote Pauli, Eddington, London, and Dirac\cite{Pauli}. However, the great
majority of theoretical physicists in the first decades of the twentieth
century remained completely unaware of Weyl's work. Therefore, it would not be
expected the general public of the time to have any possibility of accessing
Weyl's theory unless some influential and well informed science populariser
decided to write about it. It turned out that an author with that profile did
appear in scene and happened to be no one else than the renowned British
philosopher and mathematician Bertrand Russell, who was one of the best public
writer of the time.

\qquad In his book \textquotedblleft The Analysis of Matter\textquotedblright,
published in 1927, Russell dedicates an entire chapter devoted to explain and
give a critical appraisal of Weyl%
\'{}%
s theory\cite{Russel}. The book was one of the earliest and best philosophical
investigations of the physics of relativity and quantum mechanics, written by
someone who managed to master the physics, mathematics and psychology of his
time. This is particularly true with respect to theoretical physics, a subject
about what Russell had a special interest and was always aware of the latest developments.

The paper is organised as follows. In Section 2, we give a very brief account
of the essentials of Weyl's theory. We then proceed to Section 3 to review
Einstein's argument and examine in more detail the assumptions implicitly made
therein. In Section 4, we discuss Russell's appraisal of Weyl's theory and how
he saw some of the problems posed by Weyl's concerning the actual meaning of
the process of measuring time (or length) in the context of space-time
theories that still need to be clarified. We conclude with some remarks in
Section 5.

\section{\bigskip Weyl's theory briefly explained}

Weyl's theory is essentially based upon a modification of the geometry assumed
to model the space-time of the general theory of relativity, namely,
Riemannian geometry. It is perhaps one of the simplest generalization of the
latter, the only modification being that the length of a vector field may
change when parallel-transported \ along a curve in the manifold \cite{Pauli},
this change being regulated by a one-form field $\sigma\ $defined on the
space-time manifold. Let us briefly explain this new geometric property by
considering a closed curve $\alpha$ in the space-time manifold. If \ $L_{0}$
and $L$ denotes, respectively, the values of the initial and the final length
of a vector parallel-transported along $\alpha$, then \ it follows that
\[
L=L_{0}e^{\frac{1}{2}%
{\displaystyle\oint}
\sigma_{\alpha}dx^{\alpha}}.
\]
From this assumption, Weyl showed that the field $\sigma$ possesses amazing
similarities with the electromagnetic $4$-potential vector field, \ and that
was the way he found to geometrise electrodynamics.

\section{Einstein's objection}

As is well known, in an appendix to Weyl's paper, Einstein set forth a serious
objection to the theory. In his critique, Einstein argued that the theory
predicts the existence of the so-called "second clock effect" \cite{Lobo}.
According to Einstein, \ in a space-time ruled by Weyl geometry the existence
of sharp spectral lines in the presence of an electromagnetic field would not
be possible since atomic clocks would depend on their past history. Einstein
argued that this predicted effect is a logical consequence of Weyl's theory,
insofar as in a Weyl space-time the length of a vector is not held constant by
parallel transport, and this, in turn, would imply that the clock rate of
atomic clocks, measured by some periodic physical process, would be path dependent.

In order to examine Einstein's objection to Weyl's unified theory, let us
first recall two of the hypotheses on which the argument is based. They can be
stated as follows:

H1) The proper time $\triangle\tau$ measured by a clock travelling along a
curve $\alpha=\alpha(\lambda)$ is given as in general relativity, namely, by
the (Riemannian) prescription
\begin{equation}
\triangle\tau=\frac{1}{c}\int\left[  g(V,V)\right]  ^{\frac{1}{2}}%
d\lambda=\frac{1}{c}\int\left[  g_{\mu\nu}V^{\mu}V^{\nu}\right]  ^{\frac{1}%
{2}}d\lambda, \label{proper time}%
\end{equation}
where $V$ denotes the vector tangent to the clock's world line and $c$ is the
speed of light. This supposition is known as the \textit{clock hypothesis }and
clearly assumes that the proper time only depends on the instantaneous speed
of the clock and \ on the metric field.

H2) The fundamental clock rate of clocks (in particular, atomic clocks) is to
be associated with the (Riemannian) length $L=$ $\sqrt{g(\Upsilon,\Upsilon)}$
of a certain vector $\Upsilon$. As a clock moves in space-time $\Upsilon$ is
parallel-transported along its worldline from a point $P_{0}$ to a point $P$,
hence $L=L_{0}e^{\frac{1}{2}\int\sigma_{\alpha}dx^{\alpha}}$, $L_{0}$ and $L$
indicating the duration of the clock rate of the clock at $P_{0}$ and $P$, respectively.

\section{\bigskip\ Russell's comment on the problem of length measurement}

The kew point of Einstein's objection, namely, the existence of \ the second
clock effect relies entirely on the assumption that the physical process of
measuring length (or time) is well defined. This very subtil question did not
go unnoticed by Russell. In fact, in the chapter he wrote on general
relativity he starts by pointing out that he agrees with Weyl in that the
measurement of lengths of vectors at points which are not infinitesimally
close should be carried out by direct comparison. However, according to
Russell the physical process of doing this measurement is not the same of the
purely mathematical process of parallel transport of vectors. As far as we are
doing pure mathematics, this question does not appear. Nevertheless, here we
are concerned with an experimental science, and thus it seems that the
question is preliminar to any interpretation of the theory and, \textit{ipso
facto,}\ should be addressed properly. (Of course, by\textit{ length} we mean
here \textit{length in the sense of the space-time manifold}, so that it
includes the notion, for instance, of proper time and clock rates).

We then see that two assumptions are implicit in the argument used by Einstein
against Weyl's theory. The first, as we have already mentioned, refers to the
assignment of a vector the length of which gives the clock rate of a clock
carried by an observer. This assumption is also made by J. Ehlers, F. Pirani
and A. Schild (EPS) in a well known paper, in which the authors starting from
a few plausible physical hypothesis conclude that the geometry of space-time
should be given by an integrable version of Weyl geometry \cite{EPS}. The
second assumption concerns the identification of the real physical transport
of a clock with the purely mathematical operation of parallel transport of a
vector, and here we meet Russell's critical objection. At this point, we
should mention Weyl's unease with both the first and second assumptions
mentioned above. Indeed, he pointed out that the physics behind the mechanism
of a real clock was not known, and this lack of knowledge does not allow us to
accept the "geometrization" of clock rates, as we can infer from Weyl's words
\cite{Weyl}:

\textit{At first glance it might be surprising that according to the purely
close-action geometry, length transfer is non-integrable in the presence of an
electromagnetic field. Does this not clearly contradict the behaviour of rigid
bodies and clocks? The behaviour of these measurement instruments, however, is
a physical process whose course is determined by natural laws and as such has
nothing to do with the ideal process of `congruent displacement of spacetime
distance' that we employ in the mathematical construction of the spacetime
geometry. The connection between the metric field and the behaviour of rigid
rods and clocks is already very unclear in the theory of Special Relativity if
one does not restrict oneself to quasi-stationary motion. Although these
instruments play an indispensable role in praxis as indicators of the metric
field, (for this purpose, simpler processes would be preferable, for example,
the propagation of light waves), it is clearly incorrect to define the metric
field through the data that are directly obtained from these instruments.
}\cite{Weyl}

Russell starts his critical analysis of Weyl's theory by drawing a clear
distinction between a real physical transport of a clock and the mathematical
procedure of parallel transport\footnote{As Russel points out, in essence
parallel transport tries to adapt the concept of "rigid" transportation of
geometrical objects to curved spaces.}. He provides a detailed and rigorous
discussion on the problem of length measurement in general relativity, after
which he remarks that \textit{the theory raises problems which bring us
naturally to Weyl's relativistic theory of electromagnetism... }In a certain
sense, in order to avoid Einstein's objection, Weyl takes a stand regarding
Russell's comment about the need to distinguish the ideal (or mathematical)
process of congruent transference of lengths from the real behaviour of
measuring rods and clocks.

At this point, it is worth mentioning that Dirac, who revived Weyl%
\'{}%
s theory some years later, also argued that the time interval measured by
atomic clocks need not to be identified with the length of timelike vectors,
as proposed by Einstein \cite{Dirac}. In other words, it is questionable
whether clock rates are correctly modelled by the length of a certain timelike
vector, in complete accord with Russell.

However, in his critique of Weyl's theory he goes further by saying that
\textit{the theory has not been expressed with the logical purity that is to
be desired, chiefly because it is not prefaced by any clear account of what is
to be understood by "mearurement".}

\bigskip

\section{Final remarks}

From the above we see that Russell's objections are deep and preliminary to
any serious attempt to give Weyl's theory a precise physical meaning, and
therefore are at the very heart of the debate between Einstein and Weyl. And,
as we can see, all this discussion is at the root of the question whether the
second clock effect exists or not. As far as we know, neither theoretical
calculations nor any experimental attempt at measuring the magnitude of the
predicted second clock effect had been carried out until very recently
\cite{Lobo}. In fact, it is still not known by experience whether atomic
clocks, which in principle may define units of time, are integrably transported.

Incidentally, it should be pointed out that in Weyl's parallel transport the
ratio between the length of two vector transported along the same path is
constant, according to the law $L=L_{0}e^{\frac{1}{2}\int\sigma_{\alpha
}dx^{\alpha}}$). Therefore, we see that the "first clock effect" (i.e. the
twin paradox) is not affected by Weyl geometry, since the standard unit of
time follows the same variation as the total time elapsed along the path.

Finally, despite the controversy concerning its physical viability, firstly
raised by Einstein, and also the philosophical difficulties pointed out by
Russell against Weyl theory, we believe that the essential features of Weyl%
\'{}%
s theory remain untouched. As some authors have put it: "Weyl geometrical
theory contains a suggestive formalism and may still have the germs of a
future fruitful theory "\cite{Bazin}. Weyl's theory was virtually the first
robust attempt at unifying an important part of theoretical physics: gravity
and electromagnetism. It is undeniable that the quest for unification of
physics continues, now more than ever. In this respect it is worth remembering
Russell
\'{}%
s words written in his essay of 1927: \textit{It is not chimerical to hope
that a unified treatment of the whole of physics may be possible before many
years have passed }\cite{Russel}.

\bigskip

\section*{Acknowledgements}

\noindent The authors would like to thank I. Lobo for helpful discussions.
This work was partially supported by CNPq (Brazil).

\end{document}